\documentclass{article}
\usepackage{spconf,amsmath,amssymb,amsfonts,graphicx}
\newcommand*{\rom}[1]{\expandafter\@slowromancap\romannumeral #1@}

\long\def\/*#1*/{}

\title{Detecting Acoustic Reflectors using a Robot's Ego-Noise}
\name{Usama Saqib$^{\star}$, Antoine Deleforge$^{\dagger}$ and Jesper Rindom Jensen$^{\star}$}
\address{$^{\star}$Audio Analysis Lab, CREATE, Aalborg University, Denmark\\$^{\dagger}$ Universite de Lorraine, CNRS, Inria, LORIA, F-54000 Nancy, France}
\begin{document}
\ninept
\maketitle
\begin{abstract}
In this paper, we propose a method to estimate the proximity of an acoustic reflector, e.g., a wall, using ego-noise, i.e., the noise produced by the moving parts of a listening robot. This is achieved by estimating the times of arrival of acoustic echoes reflected from the surface. Simulated experiments show that the proposed non-intrusive approach is capable of accurately estimating the distance of a reflector up to 1 meter and outperforms a previously proposed intrusive approach under loud ego-noise conditions. The proposed method is helped by a probabilistic echo detector that estimates whether or not an acoustic reflector is within a short range of the robotic platform. This preliminary investigation paves the way towards a new kind of collision avoidance system that would purely rely on audio sensors rather than conventional proximity sensors.
\end{abstract}
\begin{keywords}
Robot/Drone audition, ego-noise, echo detection, robotics.
\end{keywords}
\section{Introduction}
\label{sec:intro}
Within the context of robot audition, ego-noise is defined as the noise generated by the moving parts of a robotic platform, e.g., the rotors of a drone \cite{dregon2018}. Ego-noise is a source of problems in many robotic applications, as it corrupts audio recordings captured by microphones, as available in many Human-Robot Interaction (HRI) systems \cite{nakadai2000active, okuno2015}. For this reason, ego-noise reduction is an active area of research that plays an important role in many autonomous systems, and has enabled applications such as speech recognition for HRI \cite{wake2019enhancing} or acoustic scene analysis \cite{schmidt2018}.

The structure of ego-noise has been investigated by several authors in the past. For instance, \cite{wang2016} investigates the spectral content of a multimotor aerial vehicle (MAV) and shows that the ego-noise is a combination of harmonic and broadband components. According to the authors, the noise spectra vary dynamically with the motor speed. Furthermore, because of the rigid mounting of the microphones with respect to the motors, the acoustic mixing can be assumed stationary. In \cite{schmidt2018}, the authors exploit both the spectral and the spatial characteristics of ego-noise to train a dictionary that is used for ego-noise reduction.

While robotic platforms are almost always accompanied by ego-noise, only a few studies have attempted to use it constructively in the literature. For instance, in \cite{pico2016, pico2017}, the authors emphasize that ego-noise carries useful information about the motor system's movements and the characteristics of the environments. More specifically, in \cite{pico2016}, the authors propose a forward model to predict the dynamics of the motor system of a wheeled robot. Two experiments are set to test the predictive capabilities of the model. The first experiment uses ego-noise predictions to classify velocity profiles from the auditory signals acquired by the robot, while the second experiment shows that auditory predictions can be used to detect changes in the environment, e.g., changes in the inclination of the surface where the robot is moving. Furthermore, in \cite{letizia2018}, the authors investigate the possibility of estimating a robot's motion from its ego-noise, i.e., "audio-based odometry". According to the authors, audio-based odometry presents advantages over laser- and visual-based odometry because it is not affected by changing light conditions.

In this paper, we follow this notion of using the ego-noise constructively rather than treating it as an interference, by proposing an estimator for acoustic reflectors based on the \textit{time difference of echo} (TDOE). The TDOE was introduced in \cite{diego2019} as the time difference of arrival between the direct sound source signal and its first echo in a given channel. When the source is near the receiver, the distance from the source-receiver system to the nearest acoustic reflector is half the TDOE multiplied by the speed of sound. Hence, TDOE estimation is identified to distance estimation in this paper. To estimate the TDOE, we will exploit the comb-filtering effect that emerges from the direct-path component of the sound source mixing with its delayed version, due to the presence of the acoustic reflector \cite{christensen2019introduction}. Recently, a number of methods have been proposed to use early acoustic echoes constructively for audio signal processing applications, e.g., in sound source localization \cite{diego2019}, sound source separation \cite{scheibler2018separake}, robust speaker verification \cite{al2019early} or room geometry reconstruction \cite{antonacci2012inference,dokmanic2013acoustic, crocco2014}. While the latter generalizes our study, it uses clean multi-channel room impulse responses (RIR) rather than a single noisy ego-noise signal. Accurately measuring RIRs is a time-consuming and costly process that is not suitable for robotic applications.

Conventionally, proximity sensors based on ultrasounds, lasers, or infrared lights are used to detect and localize rigid surfaces in an environment. The authors of \cite{an2018reflection} notably used laser sensors and a Kinect to estimate the positions of acoustic reflectors in a room to inform a sound source localization method. Here, we postulate that the acoustic structure of the ego-noise naturally produced by a robot carries enough information that may help in detecting and localizing acoustic reflectors solely based on audio recordings. In our previous works \cite{saqib2019, jensen2019, saqib2020estimation}, we proposed an active/intrusive approach where a loudspeaker emitting a known broadband signal was attached to a drone in order to probe the environment using times of arrival. In contrast, in this work, we propose removing the loudspeaker from the setup and develop a method solely utilizing the drone's ego-noise to detect an acoustic reflector with a single microphone. Throughout this study, we assume that the direct-path component of the ego-noise within the microphone signal is known. A number of techniques could be envisioned to estimate it, e.g., dictionary learning or model-based methods calibrated using prior measurements in an anechoic chamber, e.g., \cite{schmidt2018, saqib2019, jensen2019, saqib2020estimation}, or using close-range microphones placed next to the ego-noise sources as references. This aspect is beyond the scope of this paper and is left for future iterations of this research. Here, we focus on the specific question of whether an uncontrolled ego-noise signal, as opposed to a controlled emitted broadband signal, is sufficient to probe an environment for nearby reflectors.

First, we develop a statistically optimal TDOE estimator to solve this problem in the least-square sense. Then, we introduce a probabilistic echo-detector that helps our estimator distinguishing an acoustic reflector from empty space. Simulated experiments show that the proposed non-intrusive method is capable of accurately estimating distances of 1 meters or less and outperforms our previously proposed intrusive approach under loud ego-noise conditions. To the best of our knowledge, this is the first time that ego-noise echoes stemming from acoustic reflectors are used in a constructive way in the context of robot audition, and in particular in drone audition.

The  remainder of  this  paper  is  organized  as  follows. Section $2$ formulates the signal model and the problem. Section $3$ describes the proposed TDOE estimator based on our model. Section $4$ evaluates the performance and robustness of the proposed solution. Finally, Section $5$ concludes and provides directions for future work.

\section{Problem formulation}
\label{Sec: probFormulation}
Consider a setup with a single microphone that records both the ego-noise generated by the rotors of a drone, $x[n]$, and a background noise from the environment. The signal model is then:
\begin{align}
    y[n] &= (h*s)[n] + v[n] = x[n] + v[n],
    \label{eq: SignalModel}
\end{align}
where $h$ is the impulse response from the ego-noise source to the microphone. The source signal $s[n]$ is generated by the rotors of the drone
while $v[n]$ is the white Gaussian background noise. The signal $x[n]$ is the ego-noise of the drone that we will use to facilitate TDOE estimation. We now proceed to decompose the observed ego-noise signal, $y[n]$, as the sum of individual reflections from the source signal:
\vspace{-0.2cm}
\begin{align}
    y[n] = \sum_{q=1}^{\infty}g_{q}s[n - \tau_{q}] + v[n],
\end{align}
where $q=1$ indexes the direct path component and $q>1$ the acoustic reflections, $\tau_q$ represents the time of arrival of the $q$-th direct or reflected source signal, while $g_q$ is the corresponding gain or attenuation due to the inverse square law of sound propagation and to the sound-absorbent material at the acoustic reflector, assuming frequency-independence in this work. Acoustic impulse responses have a certain structure, which can be classified into two parts: an early part and a late part. The early part is sparse in time and contains the direct-path as well as the early reflections while the late part is characterized by a stochastic, dense and decaying tail of late reflections \cite{kuttruff2016room}. This suggests to divide the signal model as follows:
\begin{align}
\label{eq: signalModelDecomposition}
        y[n] = \sum_{q=1}^{R}g_{q}s[n - \tau_{q}] + v'[n],
\end{align}
where $R$ is the number of considered early reflections and $v'[n]$ is composed of late reflections and background noise. Furthermore, we can rewrite \eqref{eq: signalModelDecomposition} in a compact expression by separating $x[n]$ into the direct-path and early reflection components as:
\begin{align}
\label{eq: signalDirectEarlyReflection}
    y[n]=x_d[n]+x_r[n]+v'[n],
\end{align}
where $x_d[n]=g_{1}s[n-\tau_1]$ is the direct path component, and $x_r[n]=\sum_{q=2}^R g_{q}s[n-\tau_q]$ contains all the early reflections. While the direct path component $x_{d}[n]$ is always present, the reflection component $x_{r}[n]$ vanishes from the microphone signal if the robotic platform is not near any acoustic reflector. 
In drone audition applications, the microphones are often set in a fixed location with respect to the rotors. This fact could be used to estimate the direct path component using additional close-range microphones placed next to the rotors. Alternatively, a direct path estimation method calibrated in anechoic conditions could be derived. This is out of the scope of this paper, and we assume here that the direct path component $x_{d}[n]$ is known. Moreover, we are only interested in detecting one acoustic reflector, e.g., the closest one for obstacle avoidance, so we set $R = 2$ to estimate the first reflection. If we vectorize \eqref{eq: signalDirectEarlyReflection} and express it in terms of the gains and delays, we can approximate the signal model as shown:
\begin{align}
\label{eq: signalModelDirectEarlyReflection}
    \mathbf{y}[n] &\approx g_{d}\mathbf{D}_{\tau_{d}}s[n] + g_{r}\mathbf{D}_{\tau_{r}}s[n] + \mathbf{v}'[n],\\
    \mathbf{y}[n] &= \begin{bmatrix}y[n]&y[n+1]& \cdots&y[n+N-1]\end{bmatrix}^T,\notag
\end{align}
where $\mathbf{D}_{\tau}$ is a cyclic shift register that delays the unknown rotor signal $s[n]$ by $\tau$ samples and $g$ is the gain of the signal. Note that $\mathbf{D}_{\tau}$ is an identity matrix whose columns are cyclically shifted to the right by $\tau$, which approximates a true delay operator.
Similarly, we can decompose \eqref{eq: signalModelDirectEarlyReflection} into vectorized direct-path and reflection components, $\mathbf{x}_{d}[n]$ and $\mathbf{x}_{r}[n]$. Since $\mathbf{x}_{r}[n]$ is a delayed version of the direct-path component, \eqref{eq: signalDirectEarlyReflection} can be expressed as shown: 
\begin{align}
    \mathbf{y}[n] &= \mathbf{x}_{d}[n] + \frac{g_{r}}{g_{d}}\mathbf{D}_{\Delta\tau}\mathbf{x}_{d}[n] + \mathbf{v}'[n],\\\label{eq : signalModelInTermOfDirectPath}
    &=\left(\mathbf{I} + \alpha\mathbf{D}_{\Delta\tau}\right)\mathbf{x}_{d}[n] + \mathbf{v}'[n],
\end{align}
where $\Delta\tau$ is the TDOE of the observed signal, such that $\Delta\tau = \tau_{r} - \tau_{d}$ and $\alpha=\frac{g_{r}}{g_{d}}$, while $\mathbf{I}$ is the identity matrix. The task at hand is then to estimate $\Delta\widehat{\tau}$ and $\widehat{\alpha}$ in order to detect the presence of an acoustic reflector and possibly infer its distance to the nearest acoustic reflector.

\section{TDOE estimation based on Least-Squares Fit}
\label{sec:TDOE}
Let $\mathbf{y}\in\mathbb{R}^N$ contain $N$ consecutive samples of the observed signal at a given time. Assume that the corresponding direct-path component $\mathbf{x}_{d}$ is known. Then, we can estimate $\Delta\tau$ and $\alpha$ from \eqref{eq : signalModelInTermOfDirectPath} in the least-squares sense by solving the following problem:
\begin{align}
\label{eq: nlsEstimator1}
    \{\Delta\widehat{\tau}, \widehat{\alpha}\} & = \operatorname*{arg\,min}_{\Delta{\tau}, {\alpha}}
 \| \mathbf{y} - \left(\mathbf{I} + \alpha\mathbf{D}_{\Delta\tau}\right)\mathbf{x}_{d}\|^{2}\\
\label{eq: nlsEstimator}
& = \operatorname*{arg\,min}_{\Delta{\tau}, {\alpha}}J({\Delta\tau, \alpha}).
\end{align}
Note that, assuming the background noise is white and Gaussian, the resulting estimators will also be maximum likelihood estimators. The cost function in \eqref{eq: nlsEstimator} can be rewritten as follows:
\begin{align}
\label{eq: costFunction}
    J({\Delta\tau, \alpha}) &= \| \mathbf{y} - \left(\mathbf{I} + \alpha\mathbf{D}_{\Delta\tau}\right)\mathbf{x}_{d}\|^{2}\\
    &= \| \mathbf{y}\|^{2} - 2 \mathbf{y}^{T}\left(\mathbf{I} + \alpha\mathbf{D}_{\Delta\tau}\right)\mathbf{x}_{d} \notag\\
    &+ \mathbf{x}_{d}^{T}\left(\mathbf{I} + \alpha\mathbf{D}_{\Delta\tau}\right)^{T}\left(\mathbf{I} + \alpha\mathbf{D}_{\Delta\tau}\right)\mathbf{x}_{d}.\notag
\end{align}
By zeroing the derivative of \eqref{eq: costFunction} with respect to $\alpha$ we get:
\begin{align}
    \frac{\delta J}{\delta\alpha} &= -\mathbf{y}^{T}\mathbf{D}_{\Delta\tau}\mathbf{x}_{d} - \mathbf{x}_{d}^{T}\mathbf{D}_{\Delta\tau}^{T}\mathbf{y}\notag \\&+ \mathbf{x}_{d}^{T}\mathbf{D}_{\Delta\tau}^{T}\mathbf{x}_{d} + \mathbf{x}_{d}^{T}\mathbf{D}_{\Delta\tau}\mathbf{x}_{d} + 2\alpha\mathbf{x}_{d}^{T}\mathbf{D}_{\Delta\tau}^{T}\mathbf{D}_{\Delta\tau}\mathbf{x}_{d} = 0.
\end{align}
By observing that $\mathbf{D}_{\Delta\tau}^{T}\mathbf{D}_{\Delta\tau} = \mathbf{I}$, this becomes: 
\begin{align}
    -\left(\mathbf{y} - \mathbf{x}_{d} \right)^{T}\mathbf{D}_{\Delta\tau}\mathbf{x}_{d} + \mathbf{x}_{d}^{T}\mathbf{D}_{\Delta\tau}^{T}\left(\mathbf{y}-\mathbf{x}_{d}\right) + 2\alpha\|\mathbf{x}_{d}\|^{2} = 0
\end{align}
Hence,
\begin{align}
\label{eq: gainEstimator}
    \widehat{\alpha}(\Delta\tau) = \frac{\left( \mathbf{y} - \mathbf{x}_{d}\right)^{T}\mathbf{D}_{\Delta\tau}\mathbf{x}_{d}}{\|\mathbf{x}_{d}\|^{2}}.
\end{align}
We see here that the estimated gain ratio $\alpha$ has an interesting interpretation. It can be viewed as a cross-correlation between the known direct path $\mathbf{x}_{d}$ and the observed signal without direct path $\mathbf{y}-\mathbf{x}_{d}$. Now, by inserting back \eqref{eq: gainEstimator} into \eqref{eq: nlsEstimator1}, removing constant terms and simplifying, we obtain:
\begin{align}
\label{eq: tdoeEstimator}
    \Delta{\widehat{\tau}} = \operatorname*{arg\,max}_{\Delta{\tau}} \;\widehat{\alpha}(\Delta\tau)^2.
\end{align}
This expression can be maximized over a finite, predefined set of candidate delays $\Delta\tau$ to obtain the desired least-square estimate $\Delta\widehat{\tau}$.
\begin{figure}[!t]
    \centering
    \includegraphics[scale=0.5]{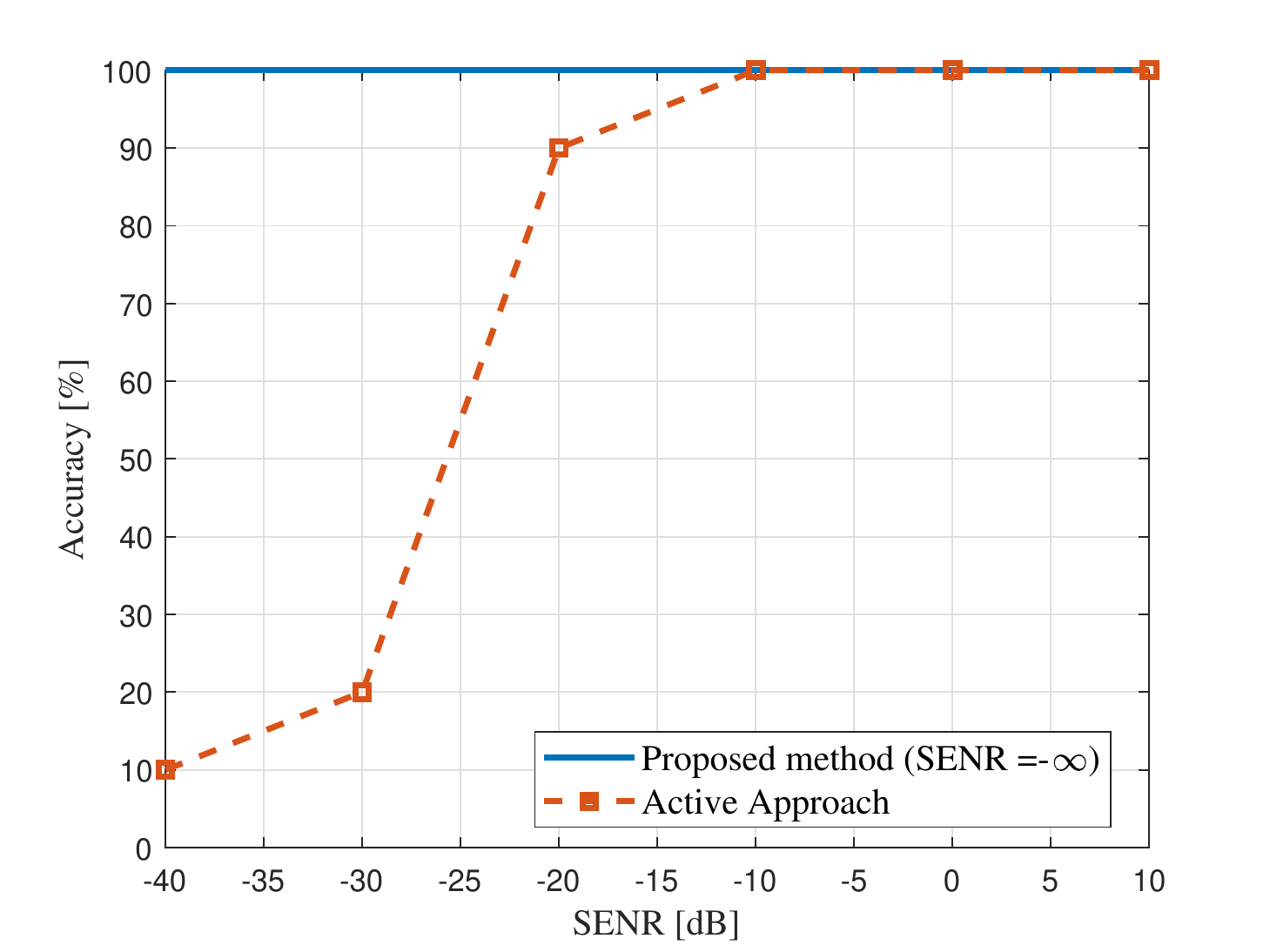}
    \caption{Comparison between the proposed TDOE estimator in \eqref{eq: tdoeEstimator} and the intrusive method in \cite{saqib2019} against varying SENRs values, when the robotic platform is  $0.5$ m from an acoustic reflector (SDNR = 40dB).}
    \label{fig: evalTdoeNlsVsSenr}
    \vspace{-0.4cm}
\end{figure}
\vspace{-0.4cm}
\section{Echo detector}
\label{sec:echoDetector}
Solving \eqref{eq: nlsEstimator1} will always give a TDOE estimate no matter where the drone is positioned in 3D space. However, we require a mechanism to distinguish whether this estimate belongs to an acoustic reflector or is an artifact stemming from the background noise. This detection is thus vital if using the TDOE estimator for, e.g., collision avoidance as it will help remove spurious estimates. Therefore, we resolve this problem by introducing an echo detector. The decision about whether an observation contains an acoustic reflection can be formulated as a detection problem \cite{kay1993fundamentals}. Let us consider the following two hypotheses:
\begin{align}
\label{eq:echoDetectorHypothesis}
    &\mathcal{H}_{0} : \mathbf{y}[n] = \mathbf{x}_{d}[n] + \mathbf{v}[n], \\
    &\mathcal{H}_{1} : \mathbf{y}[n] = \mathbf{x}_{r}[n] + \mathbf{x}_{d}[n] + \mathbf{v}[n],
\end{align}
where $\mathcal{H}_{0}$ is the null hypothesis and refers to a situation when the observation only includes the direct-path component $\mathbf{x}_{d}[n]$ and white Gaussian background noise $\mathbf{v}(n)$, with variance $\sigma^{2}$ and mean $0$, i.e., $\mathcal{N}(0,\sigma^2)$. In contrast, $\mathcal{H}_{1}$ refers to the situation when a reflected signal $\mathbf{x}_{r}[n]$ from an acoustic reflector is observed, in addition to $\mathbf{v}[n]$ and $\mathbf{x}_{d}[n]$. The observation interval is $n\in[0, N-1]$ and the generalized likelihood ratio test (GLRT) is given as:
\begin{align}
\label{echoDetectorLikelihood}
        &\mathcal{L}(n) = \frac{p(\mathbf{y};\mathbf{x}_{r}[n],\mathcal{H}_{1})}{p(\mathbf{y};\mathcal{H}_{0})}>\gamma.
\end{align}
\begin{figure}[!t]
    \centering
    \includegraphics[scale=0.5]{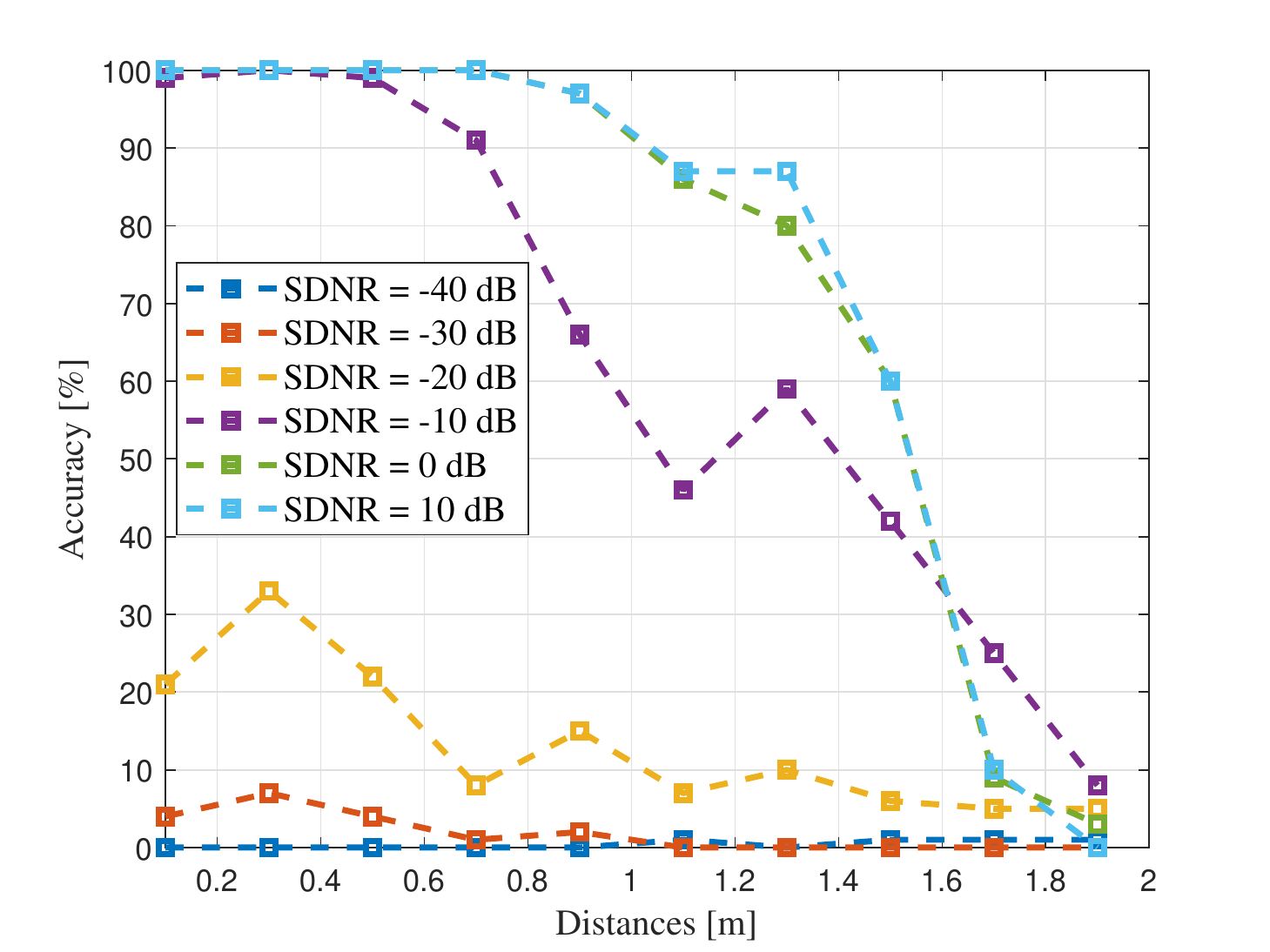}
    \caption{Evaluation of the proposed method in \eqref{eq: tdoeEstimator} against varying distances and SDNR values when the platform is near one acoustic reflectors}
    \label{fig: evalTdoaVsDistSdnr}
    \vspace{-0.4cm}
\end{figure}
In other words, in order to detect if an observation $n$ belongs to $\mathcal{H}_{1}$, its GLRT should be greater than $\gamma$.
The probability density functions (PDFs) for the two hypotheses are given as shown:
\begin{align}
    \label{echoDetectorMl}
     p(\mathbf{y};\mathbf{x}_{r}[n],\mathcal{H}_{1}) &= \frac{1}{(2\pi\sigma_{v}^{2})^{N/2}}\cdot\\&\qquad\exp{\left(\frac{-\|\mathbf{y}[n]-\mathbf{x}_{r}[n]-\mathbf{x}_{d}[n]\|^2}{2\sigma_{v}^{2}}\right)},\notag\\
    p(\mathbf{y};\mathcal{H}_{0}) &= \frac{1}{(2\pi\sigma_{v}^{2})^{N/2}}\exp{\left(\frac{-\|\mathbf{y}[n] - \mathbf{x}_{d}[n]\|^2}{2\sigma_{v}^{2}}\right)},
\end{align}
where $\sigma_{v}^{2}$ is the variance of the background noise, $v[n]$.

Note that in order to compute $\mathcal{L}(n)$, an estimate $\widehat{\mathbf{x}}_{r}[n]$ of the unknown reflected component $\mathbf{x}_{r}[n]$ is needed. One way of obtaining such estimate would be to use the estimates of section \ref{sec:TDOE}, i.e., $\widehat{\mathbf{x}}_{r}[n]=\widehat{\alpha}\mathbf{D}_{\Delta\widehat{\tau}}\mathbf{x}_d[n]$. However, this approach would strongly rely on the hypothesis that only one reflection exists in the observation. Instead, we propose to use a more straight-forward estimate for $\mathbf{x}_{r}[n]$ which is agnostic to the reflection model. Under the hypothesis $\mathcal{H}_{1}$, directly maximizing the observed-data likelihood with respect to $\mathbf{x}_{r}$ by zeroing the derivative of the logarithm of \eqref{echoDetectorMl} yields:
\begin{align}
    \label{echoDectorDifferentiation}
        &\frac{d\ln{p(\mathbf{y};\mathcal{H}_{1})}}{d \mathbf{x}_{r}} = - \left( \mathbf{x}_{r}[n] - \mathbf{y}[n] + \mathbf{x}_{d}[n]\right) = 0,
\end{align}
that is, the reflected signal is found by subtracting the direct-path component $\mathbf{x}_{d}[n]$ from the observation $\mathbf{y}[n]$, as shown:
\begin{align}
\label{echoDectorIdentity}
        \widehat{\mathbf{x}}_{r}[n] = \mathbf{y}[n] - \mathbf{x}_{d}[n].
\end{align}
By inserting \eqref{echoDectorIdentity} into \eqref{echoDetectorMl} and this back into \eqref{echoDetectorLikelihood} we get:
\begin{align}
    \label{echoDetectorThreshold}
        \ln\mathcal{L}(\mathbf{x}) &= \frac{\ln p(\mathbf{y};\mathbf{x}_{r}[n],\mathcal{H}_{1})}{\ln p(\mathbf{y};\mathcal{H}_{0})} \\&= (\mathbf{y}[n]-\mathbf{x}_{d}[n])^{T}(\mathbf{y}[n]-\mathbf{x}_{d}[n])>2\sigma_{v}^{2}\ln \gamma.
\end{align}
Hence, the criterion to detect an acoustic reflector is:
\begin{align}
    \label{thresholdValue}
    T(\mathbf{y}) &= \|\mathbf{y}[n]-\mathbf{x}_{d}[n]\|^{2}\underset{H_0}{\overset{H_1}{\gtrless}}2\sigma_{v}^{2}\ln\gamma.
\end{align}
In other words, for a reflector to be detected, the power of the reflected signal should be greater than a threshold that depends on the variance of the background noise $\sigma_{v}^{2}$. Note that this criterion will change under different noise conditions.

\section{Experiments}
\begin{figure}[!t]
    \centering
    \includegraphics[scale=0.5]{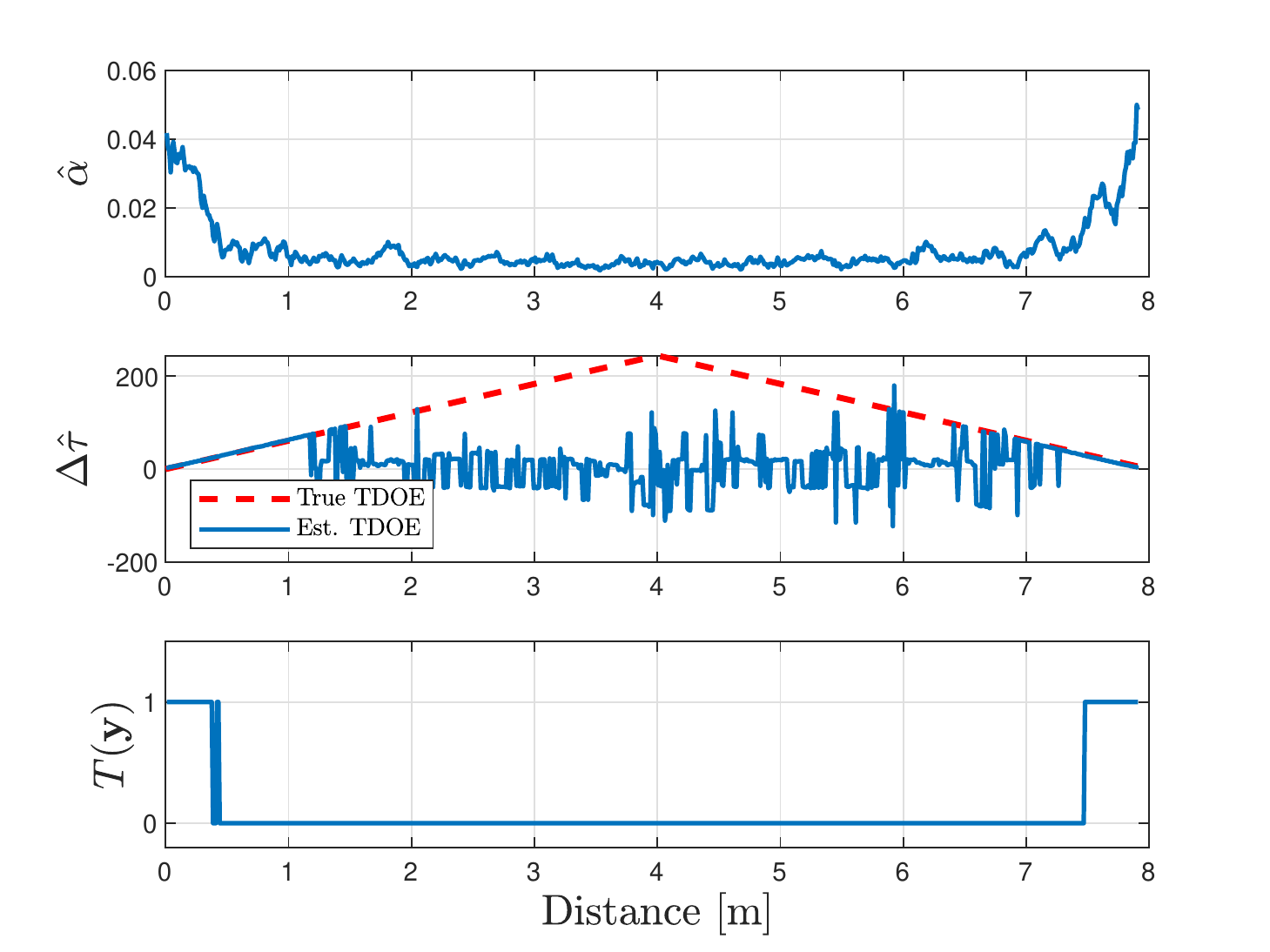}
    \caption{A moving microphone-rotor setup was tested within a simulated environment to represent a moving drone platform from one acoustic surface to another. The performance of the a) gain estimator b) TDOE estimator and c) GLRT detector is shown in the figure.}
    \label{fig:toaEstimator}
    \vspace{-0.3cm}
\end{figure}
Two experiments were conducted within a simulated room of dimensions $8\times6\times5$ m. For these experiments, we simulated the drone ego-noise as a point source. The distance between the source and the microphone is $0.2$ m. A signal generator \cite{signalGenerator2017} was used to generate the response of a moving sound source and a receiver. The signal generator convolves the sound source, i.e., the rotor noise, with a time varying RIR. The RIR is generated using the image-source method, first proposed in \cite{allen1979image}. The reverberation time was set to $T_{60}=0.4$ s, the FFT length was set to $2,048$ samples while the speed of sound was set to $343$ m/s. For the ego-noise sound source, we used the signal \texttt{allMotors\_70.wav} from the DREGON dataset \cite{dregon2018}, which was recorded from a drone whose four rotors were set to a fixed speed of 70 rotations per second. A diffuse cylindrical background noise was generated from this signal using the method described in \cite{habet2008}. The background noise has two parts, the first part is the ego-noise of the drone which contributes to late reverberation \cite{braun2018} while the latter part is the white Gaussian noise set at $40$ dB. The signal was then down-sampled from $44.1$ kHz to $5,512.5$ Hz in order to decrease the computational cost of simulating the moving source and the receiver. 
\vspace{-0.25cm}
\subsection{Comparison and evaluation of the proposed estimator}
In the first experiment, we compare the proposed method against our previously published approach \cite{saqib2019}. The latter also estimates the distance of a nearby acoustic reflector based on its TDOE, but it utilizes an embedded loudspeaker to probe the environment with a known white-noise signal $s[n]$, i.e., an \textit{intrusive} approach. Hence, in that case, the ego-noise is a disturbance in the acoustic reflector estimation. The purpose of this experiment is to test the limits of the intrusive approach under varying signal-to-ego-noise ratios (SENR). The SENR is computed in decibel (dB) as the variance of the probe signal,  $\sigma_\text{probe}^{2}$, divided by the variance of the ego-noise $\sigma_\text{ego}^{2}$. The evaluation metric used is the accuracy, defined as the percentage of TDOEs that are within $\pm10\%$ of the true TDOE, calculated from the actual distance of the robotic platform to the acoustic reflector. In this experiment, the distance to the acoustic reflector was fixed to $0.5$ m. As seen in Fig.~\ref{fig: evalTdoeNlsVsSenr}, the intrusive approach gradually fails for SENR values below $-10$ dB. For comparison, the figure also shows the accuracy obtained with the proposed approach, from the same distance, without using any probe signal (SENR=-$\infty$) and under an observed-signal to diffuse-background-noise ratio (SDNR) of 40 dB. The SDNR is calculated as the variance of the observed signal including the direct-path and reflections, $\sigma_{x}^{2}$, divided by the variance of the diffuse background environment noise $\sigma_{v}^{2}$. As can be seen, the accuracy of the proposed approach is 100\% in that case.

We then evaluate the performance of the proposed approach against varying distances and under different SDNR values. In this experiment, the robotic platform moves at a distance of $[0.1:0.2:2]$ m from the acoustic reflector, while the SDNR of the environment changes for every simulation within an interval of $[-40:10:10]$ dB. We conducted $100$ Monte-Carlo trials for each (distance, SDNR) combination to obtain the results in Fig. \ref{fig: evalTdoaVsDistSdnr}. As can be seen, the proposed TDOE estimator \eqref{eq: tdoeEstimator} provides high accuracy and offers robustness from $0$ dB and above but starts to fail under low SDNR values. Moreover, the proposed method can robustly estimate the distance of an acoustic reflector up to around $1$ m.
\vspace{-0.2cm}
\subsection{Application Example}
In Fig.\ref{fig:toaEstimator}, we simulate a scenario where we move the drone from one acoustic reflector to another, i.e., from $r_{s_{x}} = 0.1$ m to $r_{s_{x}} = 7.9$ m, in order to test the TDOE estimator in \eqref{eq: tdoeEstimator} and the echo-detector in \eqref{thresholdValue} on a larger set of distances. The value of $\ln(\gamma)$ was empirically set to $2,500$. If the echo-detector is close to an acoustic reflector, then the detector assigns a value of $1$. Otherwise, a value of $0$ is assigned to indicate empty space. The experiment was conducted under an SDNR of $10$ dB. In Fig.~\ref{fig:toaEstimator}(a), we see that the gain estimate using \eqref{eq: gainEstimator} becomes higher as the drone gets closer to an acoustic reflector. Furthermore, as seen from Fig.~\ref{fig:toaEstimator}(b), as the drone approaches an acoustic reflector, we are able to correctly estimate the TDOE up to a distance of around $1$~m. The red line on Fig. \ref{fig:toaEstimator}(b) indicates the true distance and the corresponding TDOE (ground truth) to the acoustic reflector of the drone.
However, at distances larger than $1$~m, the estimator fails to estimate the TDOE. This is illustrated by the fluctuations in the center of the figure, followed by a linear decrease of TDOE estimates as the drone approaches the other wall. Finally, results from the echo-detector using the threshold value $T(\mathbf{y})$ in \eqref{thresholdValue} are shown in Fig.~\ref{fig:toaEstimator}(c). From the figure, it is seen that the power of the reflected sound is higher than the threshold for reflectors that are closer than $0.5$ m, allowing the method to detect them. 
\vspace{-0.2cm}
\section{Conclusion and Future work}
In this paper, we proposed a TDOE estimator and an echo detector to estimate the proximity of an acoustic reflector. These make use of the natural ego-noise of a robotic platform equipped with a microphone. The proposed method could lead to the development of new sound-based collision avoidance systems for, e.g., drones. With such a system, the platform would not need to utilize proximity sensors, e.g., infrared lights or ultrasounds, to prevent collisions into walls. According to preliminary simulated experiments, the proposed method is able to estimate a distance of up to ~$1$ m and can distinguish an acoustic reflector from empty space based on the energy of the reflected signal which is compared to a predefined threshold. In future iterations of this project, we aim to investigate the estimation of the direct-path component, $\mathbf{x}_{d}[n]$, which was assumed to be known throughout this paper. This is a very challenging problem on its own, which requires further investigation. It could be addressed using the known microphone placement together with close-range microphones placed next to the ego-noise sources, or using direct-path ego-noise models trained and calibrated in anechoic conditions.



\newpage

\begin{thebibliography}{10}
	
	\bibitem{dregon2018}
	M.~Strauss, P.~Mordel, V.~Miguet, and A.~Deleforge,
	\newblock ``{DREGON}: Dataset and methods for uav-embedded sound source
	localization,''
	\newblock {\em Proc.\ IEEE Int.\ Conf.\ Intell., Robot, \ Automation.}, 2018.
	
	\bibitem{nakadai2000active}
	K.~Nakadai, T.~Lourens, H.G. Okuno, and H.~Kitano,
	\newblock ``Active audition for humanoid,''
	\newblock {\em American Asso. on Artificial Intell.}, 2000.
	
	\bibitem{okuno2015}
	H.~G. {Okuno} and K.~{Nakadai},
	\newblock ``Robot audition: Its rise and perspectives,''
	\newblock {\em Proc.\ IEEE Int.\ Conf.\ Acoust., Speech, Signal Process.},
	2015.
	
	\bibitem{wake2019enhancing}
	N.~Wake, M.~Fukumoto, H.~Takahashi, and K.~Ikeuchi,
	\newblock ``Enhancing listening capability of humanoid robot by reduction of
	stationary ego-noise,''
	\newblock {\em IEEE Transactions on Electrical and Electronic Engineering},
	vol. 14, no. 12, pp. 1815--1822, 2019.
	
	\bibitem{schmidt2018}
	A.~{Schmidt}, H.~W. {Löllmann}, and W.~{Kellermann},
	\newblock ``A novel ego-noise suppression algorithm for acoustic signal
	enhancement in autonomous systems,''
	\newblock {\em Proc.\ IEEE Int.\ Conf.\ Acoust., Speech, Signal Process.},
	2018.
	
	\bibitem{wang2016}
	L.~{Wang} and A.~{Cavallaro},
	\newblock ``Ear in the sky: Ego-noise reduction for auditory micro aerial
	vehicles,''
	\newblock {\em IEEE Int. Conf. on Adv. Video and Signal Based Surveillance},
	2016.
	
	\bibitem{pico2016}
	A.~{Pico}, G.~{Schillaci}, V.~V. {Hafner}, and B.~{Lara},
	\newblock ``How do i sound like? forward models for robot ego-noise
	prediction,''
	\newblock {\em Joint IEEE Int. Conf. on Development and Learning and Epigenetic
		Robotics}, 2016.
	
	\bibitem{pico2017}
	A.~{Pico}, G.~{Schillaci}, V.~V. {Hafner}, and B.~{Lara},
	\newblock ``On robots imitating movements through motor noise prediction,''
	\newblock {\em Joint IEEE Int. Conf. on Development and Learning and Epigenetic
		Robotics}, 2017.
	
	\bibitem{letizia2018}
	L.~Marchegiani and P.~Newman,
	\newblock ``Learning to listen to your ego-(motion): Metric motion estimation
	from auditory signals,''
	\newblock {\em Towards Autonomous Robotic Systems}, M.~Giuliani, T.~Assaf, and
	M.~E. Giannaccini, Eds., Cham, 2018.
	
	\bibitem{diego2019}
	D.~D. {Carlo}, A.~{Deleforge}, and N.~{Bertin},
	\newblock ``{MIRAGE}: {2D} source localization using microphone pair
	augmentation with echoes,''
	\newblock {\em Proc.\ IEEE Int.\ Conf.\ Acoust., Speech, Signal Process.},
	2019.
	
	\bibitem{christensen2019introduction}
	M.G. Christensen,
	\newblock {\em Introduction to Audio Processing},
	\newblock Springer International Publishing, 2019.
	
	\bibitem{scheibler2018separake}
	R.~Scheibler, D.~D. Carlo, A.~Deleforge, and I.~Dokmanic,
	\newblock ``{Separake}: Source separation with a little help from echoes,''
	\newblock {\em Proc.\ IEEE Int.\ Conf.\ Acoust., Speech, Signal Process.} IEEE,
	2018.
	
	\bibitem{al2019early}
	Khamis~A Al-Karawi and Duraid~Y Mohammed,
	\newblock ``Early reflection detection using autocorrelation to improve
	robustness of speaker verification in reverberant conditions,''
	\newblock {\em Int. J. of Speech Technology}, vol. 22, no. 4, pp. 1077--1084,
	2019.
	
	\bibitem{antonacci2012inference}
	Fabio Antonacci, Jason Filos, Mark~RP Thomas, Emanu{\"e}l~AP Habets, Augusto
	Sarti, Patrick~A Naylor, and Stefano Tubaro,
	\newblock ``Inference of room geometry from acoustic impulse responses,''
	\newblock {\em J.\ Audio, \ Speech, \ Language \ Process.}, vol. 20, no. 10,
	pp. 2683--2695, 2012.
	
	\bibitem{dokmanic2013acoustic}
	I.~Dokmani{\'c}, R.~Parhizkar, A.~Walther, Y.~Lu, and M.~Vetterli,
	\newblock ``Acoustic echoes reveal room shape,''
	\newblock {\em Proc. of the National Academy of Sciences}, vol. 110, no. 30,
	pp. 12186--12191, 2013.
	
	\bibitem{crocco2014}
	M.~{Crocco}, A.~{Trucco}, V.~{Murino}, and A.~D. {Bue},
	\newblock ``Towards fully uncalibrated room reconstruction with sound,''
	\newblock {\em Proc.\ European Signal Processing Conf.}, 2014.
	
	\bibitem{an2018reflection}
	I.~An, M.~Son, D.~Manocha, and S.~Yoon,
	\newblock ``Reflection-aware sound source localization,''
	\newblock {\em Proc.\ IEEE Int.\ Conf.\ Robotics, \ Automation.} IEEE, 2018.
	
	\bibitem{saqib2019}
	U.~Saqib and J.~R. Jensen,
	\newblock ``Sound-based distance estimation for indoor navigation in the
	presence of ego noise,''
	\newblock {\em Proc.\ European Signal Processing Conf.}, 2019.
	
	\bibitem{jensen2019}
	J.~R. Jensen, U.~Saqib, and S.~Gannot,
	\newblock ``An {EM} method for multichannel {TOA} and {DOA} estimation of
	acoustic echoes,''
	\newblock {\em Proc.\ IEEE Workshop Appl.\ of Signal Process.\ to Aud.\ and
		Acoust.}, New Paltz, NY, US, Oct. 2019.
	
	\bibitem{saqib2020estimation}
	U.~Saqib, S.~Gannot, and J.R. Jensen,
	\newblock ``Estimation of acoustic echoes using expectation-maximization
	methods,''
	\newblock {\em EURASIP J.\ on Audio, \ Speech, \ and Music Process.}, vol.
	2020, no. 1, pp. 1--15, 2020.
	
	\bibitem{kuttruff2016room}
	H.~Kuttruff,
	\newblock {\em Room acoustics},
	\newblock Crc Press, 2016.
	
	\bibitem{kay1993fundamentals}
	S.~M. Kay,
	\newblock {\em Fundamentals of statistical signal processing}, vol.~2,
	\newblock Prentice Hall PTR, 1993.
	
	\bibitem{signalGenerator2017}
	E.~A.~P. Habets,
	\newblock ``Signal generator,''
	\newblock Tech. {R}ep., Friedrich-Alexander-Universität Erlangen-Nürnberg
	(FAU), 2017.
	
	\bibitem{allen1979image}
	J.~B. Allen and D.~A. Berkley,
	\newblock ``Image method for efficiently simulating small-room acoustics,''
	\newblock {\em J.\ Acoust.\ Soc.\ Am.}, vol. 65, no. 4, pp. 943--950, 1979.
	
	\bibitem{habet2008}
	E.~A.~P. Habets, I.~Cohen, and S.~Gannot,
	\newblock ``Generating nonstationary multisensor signals under a spatial
	coherence constraint,''
	\newblock {\em J.\ Acoust.\ Soc.\ Am.}, vol. 124, no. 5, pp. 2911--2917, 2008.
	
	\bibitem{braun2018}
	S.~{Braun}, A.~{Kuklasiński}, O.~{Schwartz}, O.~{Thiergart}, E.~A.~P.
	{Habets}, S.~{Gannot}, S.~{Doclo}, and J.~{Jensen},
	\newblock ``Evaluation and comparison of late reverberation power spectral
	density estimators,''
	\newblock {\em J.\ Audio, \ Speech, \ Language \ Process.}, vol. 26, no. 6, pp.
	1056--1071, 2018.
	
\end{thebibliography}

\end{document}